\documentclass[5p,number,sort&compress]{elsarticle}

\usepackage{lineno}
\modulolinenumbers[5]
\usepackage[colorlinks=true, allcolors=blue]{hyperref}% add hypertext capabilities
\usepackage{braket}
\usepackage{amsmath}
\usepackage{amssymb}
\usepackage{siunitx}
\journal{arXiv}

\begin{document}

\def \figureOne {
\begin{figure}[!t]
\includegraphics[scale=1]{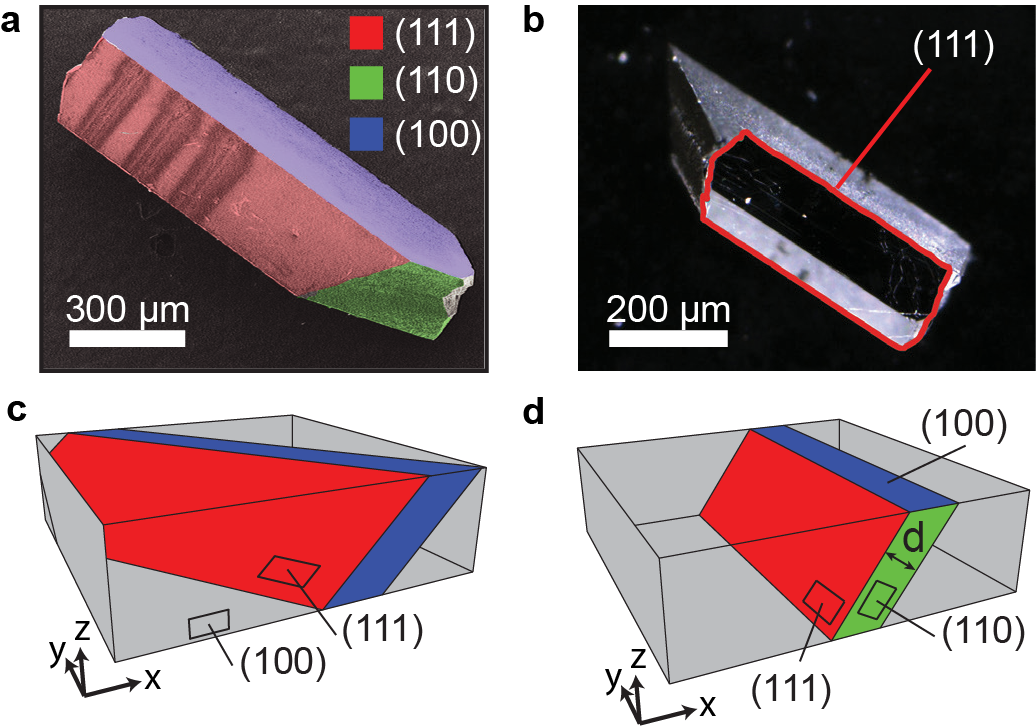}
\caption{\label{optical_image} \textbf{Laser-nucleated cleaving of single crystal diamond.} (a) Scanning electron micrograph and (b) bright-field optical image of (111)-faced plates produced from a standard grade (SG) (100)-faced single-crystal diamond. Illustration of the orientation of a (111)-faced plate that can be produced from (c) an SG (100)-faced sample with (100) sides and (d) an electronics grade (EG) sample with (110) sides.}
\end{figure}
}

\def \cleaveProcess{
\begin{figure*}[!t]
\includegraphics[scale=1]{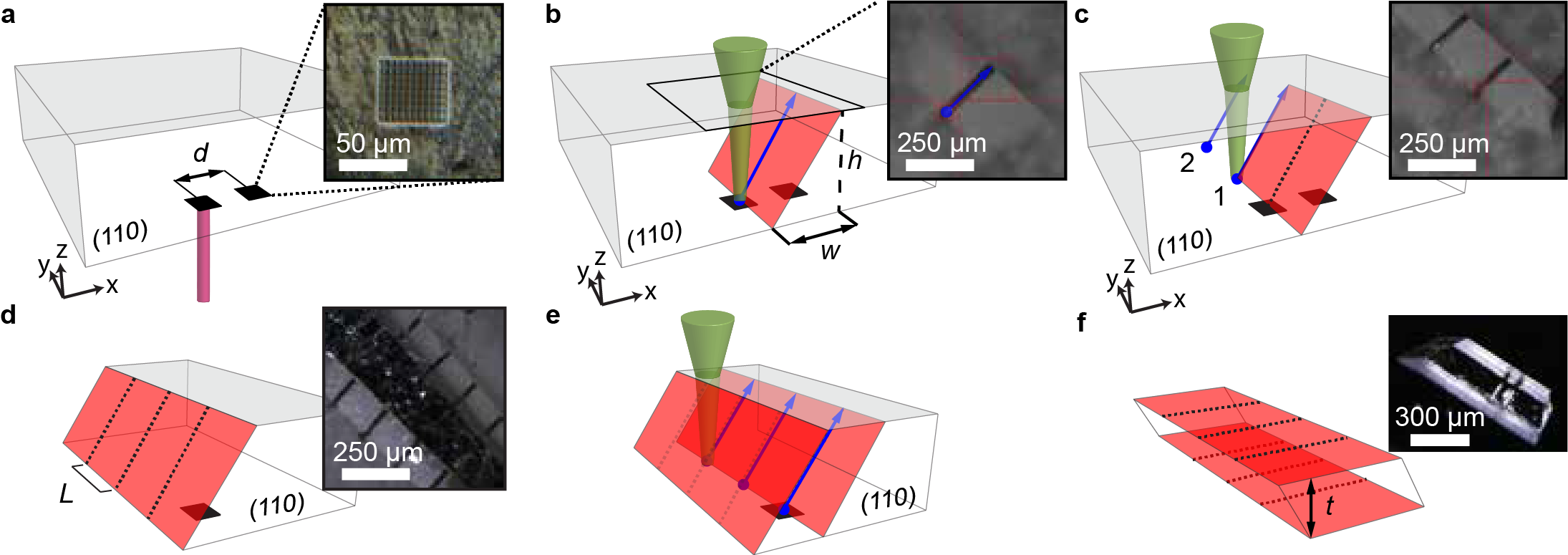}
\caption{\label{cleave_process}\textbf{LNC process overview.} Laser-nucleated cleaving process for an EG-SCD starting plate with (100) top and bottom faces and (110) side walls. (a) Fabrication of textured nucleation sites produced with a UV Excimer laser. (b) A (111) cleavage plane is generated by focusing the laser below the bottom surface and rastering away from a nucleation site in the $x$-direction. (c) The (111) cleavage plane is extended by repeating the laser raster at discrete steps in $y$. (d) Illustration of the complete cleave for a single (111) face. (e),(f) The process is repeated for an adjacent nucleation site in order to cleave a plate with parallel (111) faces.} 
\end{figure*}
}
 
\def\samples{
\begin{figure}[!t]
\includegraphics[scale=1]{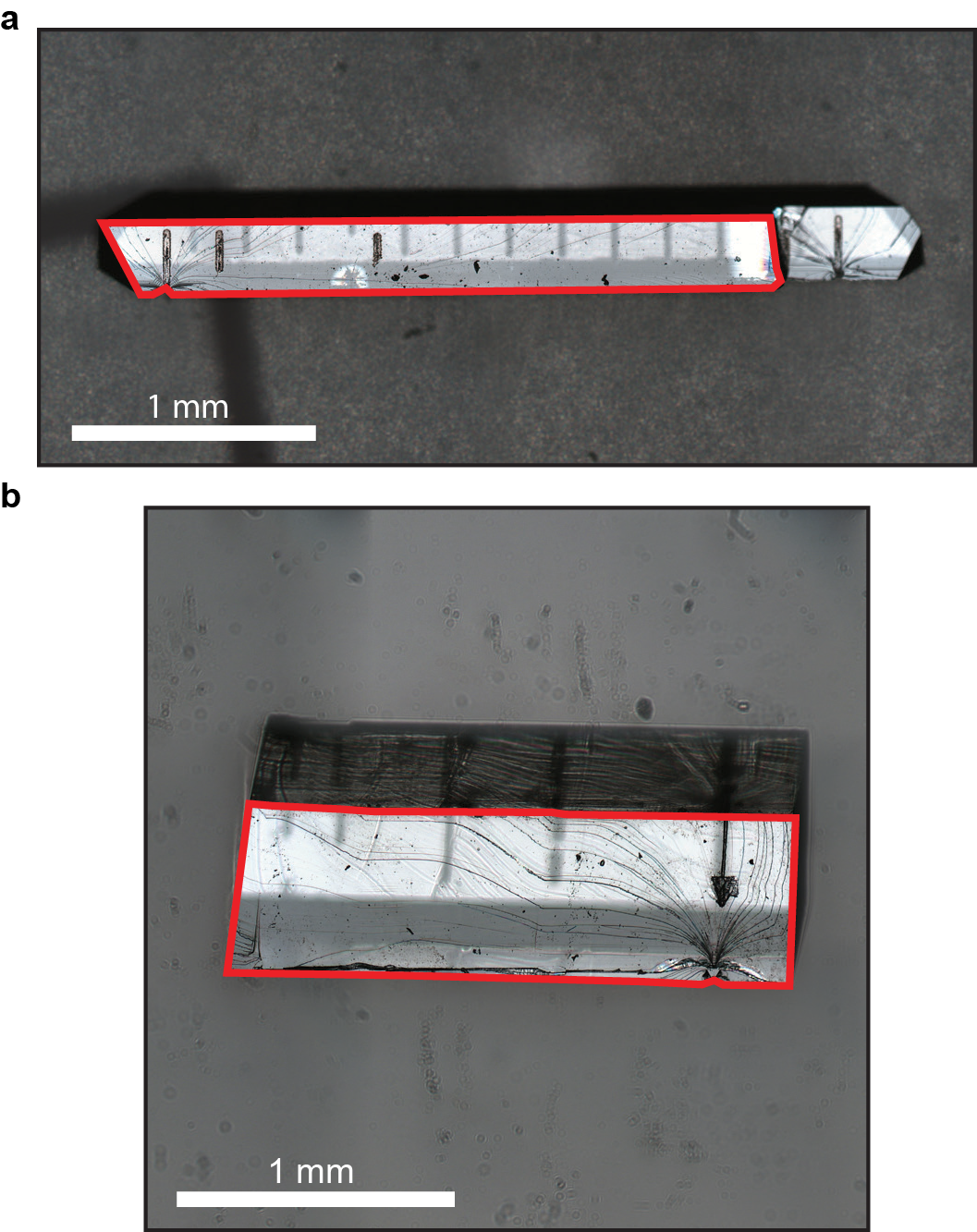}
\caption{\label{samples} \textbf{SCD (111)-faced plates fabricated by LNC}. Bright-field microscope images of (a) a \SI{3}{\milli\meter} $\times$ \SI{0.3}{\milli\meter} $\times$ \SI{0.17}{\milli\meter}-thick SG-SCD plate and (b) a \SI{2.0}{\milli\meter} $\times$ \SI{0.6}{\milli\meter} $\times$ \SI{0.52}{\milli\meter}-thick EG-SCD plate.  Top (111)-faces are outlined in red.}
\end{figure}
}

\def\diamondsurface{
\begin{figure}[!t]
\includegraphics[scale=1]{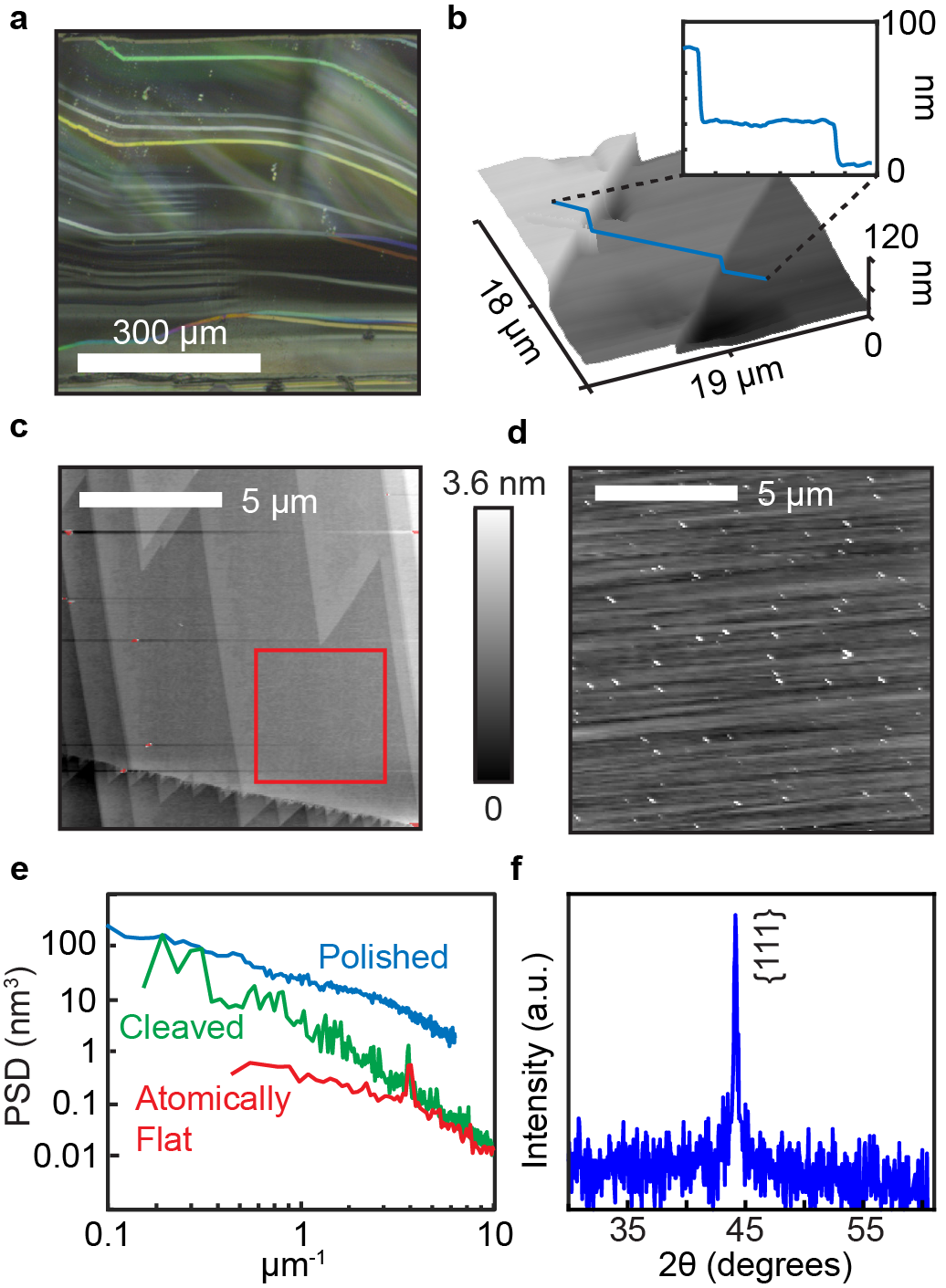}
\caption{\label{diamond_surface} \textbf{Surface morphology.} (a) Differential-interference contrast image of an LNC, EG-SCD plate showing Wallner lines on the cleaved surface. (b) Atomic-force micrograph area scan and profile line-scan inset of Wallner lines. (c) Atomic-force micrograph of atomic step edges found in the area between Wallner lines. (d) Atomic-force micrograph of the as-received EG-SCD polished surface. (e) 1D power spectra extracted from the the AFM micrographs in (c) and (d) as well as the cropped region on an atomically-flat terrace in (c). (f) X-ray diffraction curve verifying that the face of the plate is (111)-oriented.}
\end{figure}
}

\def \fluence {
\begin{figure}[!t]
\includegraphics[scale=1]{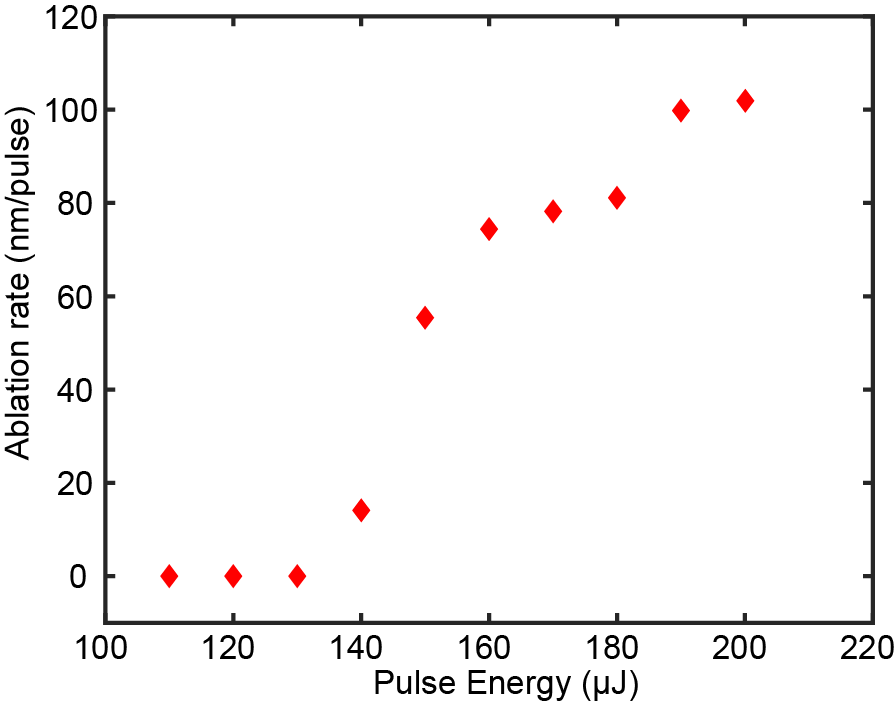}
\caption{\label{fluence} \textbf{Ablation rate.} Measured ablation rate as a function of laser pulse energy, when focused at the diamond surface.}
\end{figure}
}

\def \Wallners {
\begin{figure}[!t]
\includegraphics[scale=1]{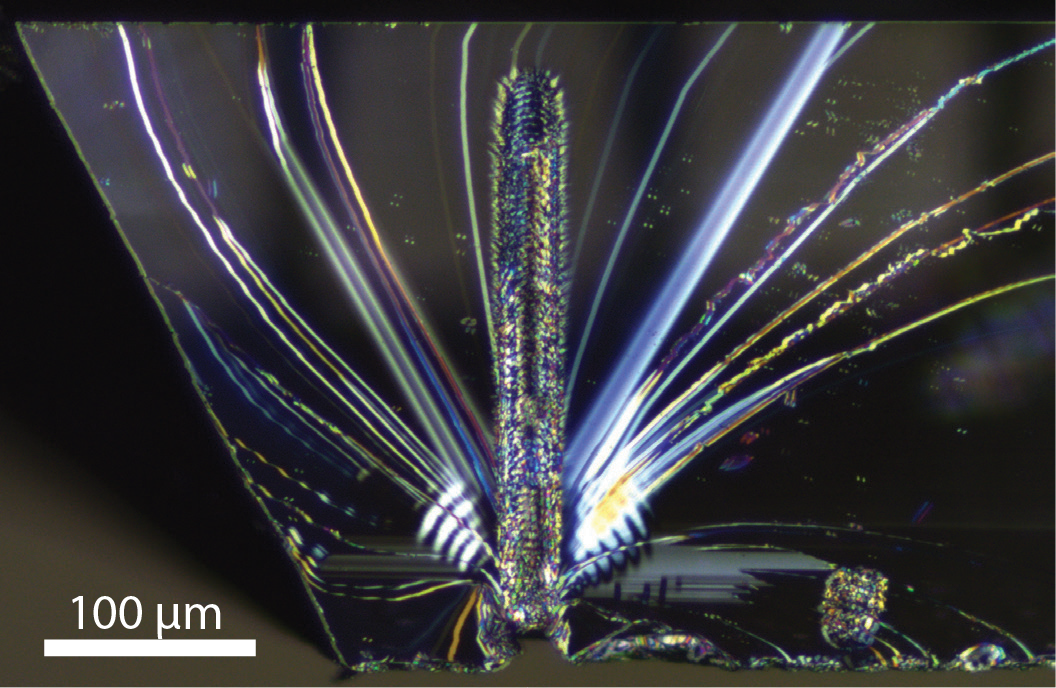}
\caption{\label{Wallners} \textbf{Cleave nucleation}. Circularly-polarized differential interference contrast (CDIC) image of the first cleave line on an SG-SCD (111) face, illustrating how Wallner lines emanate from the laser line.}
\end{figure}
}

\begin{frontmatter}

\title{Fabrication of (111)-Faced Single-Crystal Diamond Plates\\ by Laser Nucleated Cleaving}

%% Group authors per affiliation:
\author[qel]{Samuel M. Parks}
\author[qel]{Richard R. Grote}
\author[qel,dop]{David A. Hopper}
\author[qel]{Lee C. Bassett\corref{cor}}
\ead{lbassett@seas.upenn.edu}
\address[qel]{Quantum Engineering Laboratory, Department of Electrical \& Systems Engineering,\\ University of Pennsylvania, Philadelphia, PA 19104, United States}
\address[dop]{Department of Physics, University of Pennsylvania, Philadelphia, PA 19104, United States}
\cortext[cor]{Corresponding author.}

\begin{abstract}
Single-crystal diamond plates with surfaces oriented in a (111) crystal plane are required for high-performance solid-state device platforms ranging from power electronics to quantum information processing architectures. However, producing plates with this orientation has proven challenging. In this paper, we demonstrate a method for reliably and precisely fabricating (111)-faced plates from commercially available, chemical-vapor-deposition-grown, type-IIa single-crystal diamond substrates with (100) faces. Our method uses a nanosecond-pulsed visible laser to nucleate and propagate a mechanical cleave in a chosen (111) crystal plane, resulting in faces as large as $\SI{3.0}{\milli\meter}\times\SI{0.3}{\milli\meter}$ with atomically flat surfaces, negligible miscut angles, and near zero kerf loss. We discuss the underlying physical mechanisms of the process along with potential improvements that will enable the production of millimeter-scale (111)-faced single-crystal diamond plates for a variety of emerging devices and applications.
\end{abstract}


\end{frontmatter}

%\linenumbers

\section{Introduction}
 
 % Why is single-crystal diamond important?
Advances in homoepitaxial chemical vapor deposition (CVD) of high-purity single-crystal diamond (SCD) have made the exceptional material properties of SCD available for a variety of new and exciting applications \cite{Balmer_JPCM_09,Gracio_JPD_10,E6_handbook_17,Tokuda_NAD_15}. In particular, the wide bandgap, high carrier mobility, large thermal conductivity, corrosion resistance, and biocompatibility of SCD have enabled new devices for high-power electronics \cite{Achard_DRM_11}, ultraviolet light sources \cite{Makino_APL_11} and detectors \cite{Balducci_APL_05}, nonlinear optics \cite{Mildren_OpEx_08, Hausmann_NP_14, Latawiec_Optica_15}, quantum information processing \cite{Toyli2013}, biomedical applications \cite{Balasubramanian_COBC_14, Narayan_MT_11}, magnetometry \cite{Rondin_RPP_14}, and integrated photonics \cite{Aharonovich_NP_11, Aharonovich_AOM_14, Schroder_JOSAB_16,Grote_APLP_16,Gao_JLT_16}. 

\figureOne
\cleaveProcess

% Why are {111} faces important?
Many of these diverse applications benefit from the advantageous mechanical, thermal, and bond properties along $\langle 111\rangle$. For example, the (111) crystal plane is the hardest face \cite{Wilks1972,Field_RPP_12}, making it more resilient to damage and desirable for industrial scale diamond cutting tools \cite{Balmer_JPCM_09}. Similarly, the Raman gain coefficient is strongest when the optical polarization axis is aligned along $\langle 111\rangle$ \cite{Mildren_OED_13,Sabella_OL_10}. For this reason, integrating on-chip Raman lasers requires precise crystallographic orientation of the diamond layer to maximize the laser efficiency~\cite{Latawiec_Optica_15}.  High-power electronic devices also benefit from (111) faces, which provide improved donor incorporation efficiency and correspondingly higher mobility \cite{Katagiri_APL_04,Kato_APL_05,Kato_DRM_11,Kato_APL_16,Achard_DRM_11,Tallaire_DRM_14}. Finally, point defects in SCD that are used as single-photon sources and spin qubits, such as the nitrogen-vacancy and silicon-vacancy centers, have a symmetry axis aligned along $\langle 111\rangle$ directions \cite{Awschalom2013,Hepp_PRL_14,Bhaskar_arXiv_16,Neu_APL_14}. Thus, a (111) face maximizes the interaction with normally-incident light fields and optimizes the photon collection efficiency from these emitters, which is critically important for light-matter quantum interfaces \cite{Bernien2013} and high-sensitivity magnetometry \cite{Pham_PRB_12,Ozawa_APE_17}.

% State of the art in making {111} plates, and how our approach differs and what the advantages are
While (100)-faced SCD plates are commonly produced by homoepitaxial CVD, growth of (111)-faced SCD plates has proven challenging due to twinning and stacking faults that form along $\langle 111\rangle$ directions \cite{Friel_DRM_09, Gracio_JPD_10,Tallaire_DRM_14}. Despite some recent advances \cite{Tokuda_NAD_15,Widmann_DRM_16}, inherent challenges still remain in growing (111)-oriented layers thicker than $\approx$\SI{100}{\nano\meter} with low surface roughness. For this reason, macroscopic (111)-faced plates are typically cut from larger (100)-faced starting plates using micro-machining laser ablation tools \cite{Stoupin2016} and subsequent polishing steps are used to reduce the surface roughness \cite{Hird_OED_13,Konov_OED_13}.  This process results in substantial loss of costly SCD material (kerf loss) along with misalignment of the resulting face relative to [111], since it is nearly impossible to polish the (111) face using traditional methods \cite{Wilks1972}.  Furthermore, the mechanical polishing step induces sub-surface damage that can only be removed by further chemical etching \cite{Friel_DRM_09}. 

An alternate approach, traditionally used to shape diamond gemstones, is mechanical cleaving, where a small scribe, or kerf, is created and struck by hand with a knife to initiate a cleave. It is not feasible, however, to perform this process by hand with the micron-scale accuracy needed to create large, thin, parallel plates for electronic and optical applications from small samples of SCD.  Instead, it would be valuable to establish a process similar to silicon stealth dicing \cite{Kumagai_TSM_07}, whereby a laser is used to nucleate and propagate a mechanical cleave with high spatial precision and control of the cleaving dynamics.  

% what we do here
% either move earlier or later with summary paragraph
In this paper, we present a laser-nucleated cleaving (LNC) process for SCD.   Rather than machining the entire surface of a desired face, the LNC process uses a pulsed, focused laser with sub-bandgap photon energy (\SI{532}{\nano\meter}) to locally graphetize a thin ($\approx$\SI{30}{\micro\meter} diameter) line that propagates a mechanical cleave along a desired (111) plane.  As shown in Fig.~\ref{optical_image}, (111)-faced plates are fabricated by generating two parallel (111) cleave planes, forming a parallelepiped shape. This process results in large, atomically-flat (111)-faced SCD plates while eliminating kerf loss and removing the need for mechanical polishing.  

% Move the following section to the discussion
% The LNC process has potential for reducing or completely eliminating plate wedge, which adversely affects diamond waveguide-based devices \cite{Latawiec_Optica_15,Hausmann_NP_14,Hausmann_NL_13,Gao_JLT_16}. Furthermore, LNC causes minimal kerf loss allowing for the entire (100)-faced starting plate to be divided into the maximum number of (111)-faced plates. LNC may also prove useful for fabricating (111)-faced high-pressure/high-temperature (HPHT)-grown seed crystals required for [111] CVD growth \cite{Friel_DRM_09, Gracio_JPD_10,Tallaire_DRM_14,Tokuda_NAD_15,Widmann_DRM_16}.

% transition to rest of paper
The paper is organized as follows: first, a step-by-step description of the LNC process is provided, followed by atomic force microscopy (AFM) and X-ray diffraction measurements that determine the surface roughness and confirm the crystal orientation, respectively. Then, the potential physical mechanisms that contribute to the LNC process are discussed. Finally, avenues for improving the process to optimize the size and flatness of (111)-faced SCD plates are presented.

% These improvements are have the potential to enable the reliable production of atomically-flat, thin (111)-faced SCD plates for applications in cavity quantum electrodynamics \cite{Janitz_PRA_15} and diamond integrated optics \cite{Hausmann_NL_13, Gao_JLT_16} with minimal kerf loss and wedge.

\section{Materials and methods} 
% overview of process and description of Fig 2
The LNC process is illustrated in Fig.~\ref{cleave_process} and the individual steps are described in the section below. We begin by discussing the types of SCD plates used in this study followed by the micro-machining laser specifications. Then, we outline the LNC process which is grouped into four sub-processes consisting of nucleation site creation, initiation of the cleave, propagating the cleave front, and backside cleaving to produce a free standing (111)-faced SCD plate. Finally, the details concerning the characterization of the SCD plates' physical properties are covered. Additional details are included in the supporting information \cite{Supporting}.

\subsection{Materials}
We perform LNC on commercially available  ``standard grade'' (SG, $<1$~ppm nitrogen and boron) and high-purity ``electronics grade'' (EG, $<5$~ppb nitrogen and boron) type-IIa, CVD single-crystal diamond from Element6 \cite{E6_website}.  The edge and face orientations of these samples are shown in Fig.~\ref{optical_image}(c)-(d). 

\subsection{Nucleation site creation}
The LNC process begins with the creation of a nucleation site on the bottom surface of the starting SCD plate. The nucleation site acts as a seed for producing the initial cleave.  For samples with two polished (100) faces, textured regions [Fig.~\ref{cleave_process}(a), inset] are laser-machined onto the backside using an excimer laser \cite{Supporting}. Only one textured site per desired (111) face is required. For single-side polished pieces, the unpolished side of as-received samples with a surface roughness of $R_{q} = ~\SI{250}{\nano\meter}$ \cite{E6_website} was found to be sufficient to nucleate cleaves without any additional patterning.

\subsection{Micromachining laser beam preparation}
We perform LNC using a nanosecond pulsed, frequency-doubled Yb-doped fiber laser operating at a wavelength of \SI{532}{\nano\meter} with a repetition rate of \SI{150}{\kilo\hertz}, integrated in a micromachining laser system (IPG Photonics, IX-280-ML). The primary focusing optic is a \SI{100}{\milli\meter}-focal-length objective that produces a beam waist of diameter $<$\SI{17}{\micro\meter}, a beam quality factor of $M^2 = 1.5$, and a depth of field $\approx${\SI{800}{\micro\meter}. In this system, the laser focus and course lateral position are controlled using a motorized stage, and cuts are made by steering the beam with a galvo scanner. The starting SCD plate is fixed to a \SI{1.0}{\milli\meter}-thick glass microscope slide with double-sided tape so that the micromaching laser can be focused through the air/diamond interface (referred to as the top surface) onto the tape-diamond interface (referred to as the bottom surface).  

The beam focus is positioned such that the bottom surface of the SCD plate ablates while the top surface does not. SCD ablation by sub-bandgap light is a nonlinear optical process, and the ablation rate depends sensitively on the pulse fluence \cite{Konov_LPR_12}.  Positioning the laser focus close to the bottom surface establishes a fluence gradient such that the top surface is  unaffected by the micromaching laser beam, whereas the bottom surface is above the threshold for ablation and graphetization. The presence of a nucleation site or unpolished bottom surface further lowers the threshold for ablation. Figure~\ref{fluence} depicts the experimentally measured ablation rate of a top polished SCD surface as a function of pulse energy for our micromaching laser, exhibiting an abrupt threshold around \SI{140}{\micro\joule}.

%To properly set the fluence gradient for our system, we start by moving the surface of the starting plate to the $z$ position of the laser beam focus. Next, the sample is translated parallel to the beam (along $z$) by \SI{415}{\micro\meter} such that the laser is focused through the starting plate to a point below the bottom surface. Vertical dashed lines in Fig.~\ref{fluence} indicate the resulting laser fluences at the top surface, the bottom surfaces of a \SI{300}{\micro\meter}-thick SG plate and a \SI{500}{\micro\meter}-thick EG plate, and at the laser focus, assuming a Gaussian beam. We calculate that this procedure results in a 19$\%$ difference in intensity at the top surface as compared to the bottom surface of a \SI{300}{\micro\meter}-thick sample, and 32$\%$ difference for \SI{500}{\micro\meter}-thick sample. While modest, this intensity difference is all that is required due to the highly nonlinear nature of the ablation mechanism \cite{Konov_LPR_12}.

\fluence

\subsection{Initiating the cleave}
To initiate a cleave, the micromachining laser beam is aligned over a textured nucleation site, or a target region of an unpolished bottom surface using the sample stage. The laser beam is then switched on and laterally scanned in $x$ at a speed of \SI{8}{\centi\meter\per\second} by a distance $w$, as illustrated in Fig.~\ref{cleave_process}(b). The scan distance is calculated according to $w=h/\tan\SI{55}{\degree}$, where $h$ is the starting SCD plate thickness, and \SI{55}{\degree} is the angle between the [111] and [100] lattice vectors. This translation distance ensures that the cleave extends through the full thickness of the SCD ($x$-$z$ plane). After this step, a darkened line appears and a new reflective surface can be seen within the diamond [Fig.~\ref{cleave_process}(b), inset]. The combination of these two observations heralds the presence of a cleave along a (111) face. 

\subsection{Propagating the cleave}
As seen in Fig.~\ref{cleave_process}(b), a typical cleave only propagates a few hundred microns away from the laser scan line in the $y$-direction. The (111) crystal face is extended by repeating the lateral laser scans used in the previous step, but with the beam translated along $y$ near the edge of the newly-created (111) face. Scanning the laser again along $x$ extends the cleave further along $y$, as shown in Fig.~\ref{cleave_process}(c).  As such, a textured nucleation site is only required at the position of the first laser scan line to initiate a cleavage plane. After each subsequent extension of the (111) face, a new boundary is identified in the microscope and the laser scan process is repeated until the cleave has been extended across the entire width of the starting SCD plate, creating a fully exposed (111) face [Fig.~\ref{cleave_process}(d)].  

%The distance $L$ could then be hard-coded into a turn-key procedural script running the micromachining laser for performing a complete cleave across the sample with no manual adjustment other than the initial aiming of the beam necessary.

% This is a discussion worthy sentence
%Further optimization of this reduced ablation line technique would result in larger usable \{111\} plates.

%Maybe move the following discussion of dry etch to remove the graphite destroyed layers to the discussion?
%These parameters are related by $c=h/\tan\SI{55}{\degree}$. $L$ was chosen in an attempt to minimize the amount of ablation damage to the (111)-face after cleaving but also ensure that cleave propagation [Fig~\ref{cleave_process}(c)] occurred with as few unwanted cracks as possible, although this can potentially be removed by a chlorine-based dry etch \cite{Lee_DRM_08}. Our exploration of this process used a value of $L=\SI{300}{\micro\meter}$, however no rigorous attempt was made to optimize this value.  

\subsection{Backside cleaving}
To complete the (111)-faced SCD plate fabrication, a second (111) crystal face is generated that is parallel to the initial exposed face. The distance between the two nucleation sites sets the resulting plate thickness $t$ as can be seen in Fig.~\ref{cleave_process}(f). For reliable performance, we find that the second nucleation site should be offset by a distance greater than the beam translation distance, ($d>w$) [Fig.~\ref{cleave_process}(e)], hence the minimum (111)-faced plate thickness is limited by the starting plate height, $h$. 

\subsection{Characterization}
% XRD
Measurements of the cleaved-face lattice orientation were made at room temperature using X-Ray diffraction with a $\lambda = \SI{0.1544}{\nano\meter}$ source, a \SI{1}{\degree} divergence slit, \SI{1}{\degree} scatter slit, \SI{0.6}{\milli\meter} receiving slit, and a scintillator detector.  Plates were mounted on a glass slide using crystal bond.
% AFM
Surface roughness was characterized using an Asylum MFP-3D atomic-force scanning probe with $z$-axis noise $<$\SI{0.06}{\nano\meter} (average deviation) in a \SI{0.1}{\hertz}-\SI{1}{\kilo\hertz} bandwidth.  

\section{Results}
\samples
% Intro
%The results of our LNC process are presented through optical images of our (111)-faced SCD plates that show the relative size and quality that can be achieved. Characterization of the surface roughness over a wide range of lateral length scales is then presented. Finally, the crystallographic orientation is confirmed with X-Ray diffraction measurements.\note[LB]{These intro sentences seem unnecessary to me. You could just start with the next paragraph. However, what about adding subheadings like ``Face areas'', ``Roughness'', ``Face orientation''?} 

% Examples of plates, summary of areas, discussion of kerf loss
\subsection{Plate dimensions}
Representative (111)-faced plates produced by the LNC process are shown in Fig.~\ref{optical_image}(a)-(b), Fig.~\ref{cleave_process}(f), and Fig.~\ref{samples}. The dark lines on the top and bottom surfaces of the plates are the locations of the micromachining laser scan lines. A laser-scan-line separation in $y$ of $L=$~\SI{300}{\micro\meter} was found to be suitable for reliably propagating the cleave using an automated process [Fig.~\ref{samples}, bottom surface]. However, by manually identifying the cleavage plane boundary in the microscope, much larger spacings can be realized [Fig.~\ref{samples}, top surface], resulting in laser-scan-line-free areas ranging from \SI{0.18}{\milli\meter}$^2$ [Fig.~\ref{cleave_process}(f) inset] to \SI{1.2}{\milli\meter}$^2$ [Fig.~\ref{samples}(b)]. 

Fabricated plate thicknesses vary from \SI{50}{\micro\meter} [Fig.~\ref{cleave_process}(f) inset] to \SI{520}{\micro\meter} [Fig.~\ref{samples}(b)]. Minimum thicknesses of \SI{170}{\micro\meter} and \SI{290}{\micro\meter} are reliably produced for starting plate thicknesses of \SI{300}{\micro\meter} and \SI{500}{\micro\meter}, respectively. A single $\SI{3}{\milli\meter} \times \SI{3}{\milli\meter}$ (100) SCD plate can typically yield up to 10 (111)-faced plates. The number is limited due to the increasing number of stochastically-generated cracks along undesired planes as the original SCD plate is diced into multiple (111)-faced plates. 

\diamondsurface

% Surface roughness (Wallner -> atomic step edges)
\subsection{Surface Morphology}
Lines running along the (111) face in the direction of the cleave propagation can be observed in the differential-interference-contrast optical microscope image shown in Fig.~\ref{diamond_surface}(a). These nanometer-scale steps or ridges are known as Wallner lines, which appear ubiquitously on cleaved surfaces of brittle materials as a result of the interaction between the crack front and transverse acoustic waves \cite{Rabinovitch_JAP_06,Hird_OED_13}. Using atomic force microscopy (AFM), we find that the Wallner lines typically exhibit step heights of 10--\SI{50}{\nano\meter} [Fig~\ref{diamond_surface}(b)], with separations that vary from $\approx$\SI{10}{\micro\meter} near the initial laser scan line to $>$\SI{100}{\micro\meter} in regions farther away. 

As shown in Fig.~\ref{diamond_surface}(c), the areas between Wallner lines exhibit atomically-flat (111) surfaces characterized by micron-scale, triangular terraces with step heights consistent with the expected \SI{0.206}{\nano\meter} spacing of (111) atomic planes \cite{Tokuda_NAD_15}. The terraces are occasionally intersected by linear features as observed in the bottom-left of Fig.~\ref{diamond_surface}(c). For comparison, the polished (100) surface of an EG-SCD plate, as-received from the manufacturer \cite{E6_website} following cleaning using an acid bath and O$_2$ plasma treatment, is displayed in Fig.~\ref{diamond_surface}(d). 

Clearly, the morphologies of the cleaved and polished surfaces are very different.  Figure~\ref{diamond_surface}(e) shows the power spectral density (PSD) of the AFM scans in Figs.~\ref{diamond_surface}(c,d), as well as the PSD for the region of a single atomic plateau  indicated by a red box in Fig.~\ref{diamond_surface}(c). Over the entire $\SI{12}{\micro\meter}\times\SI{12}{\micro\meter}$ AFM scan windows, we calculate root-mean-square (RMS) surface roughness values of $R_{q}=\SI{0.22}{\nano\meter}$ and \SI{0.46}{\nano\meter} for the cleaved (111) and polished (100) surfaces, respectively.  The LNC surface roughness represents an order-of-magnitude improvement compared to the manufacturer's specification for the polished surface (average roughness $<$\SI{5}{\nano\meter}) and a smoothness comparable to the best values reported from advanced polishing techniques  of $<\SI{0.1}{\nano\meter}$ \cite{DDK}. Furthermore, the majority of the cleaved-surface roughness occurs in the form of atomic steps with micron-scale spacings, whereas the polished surface exhibits roughness over a broad range of spatial frequencies.  The calculated RMS roughness of the \SI{25}{\micro\meter\squared} region on an atomic terrace is $R_{q} = \SI{0.05}{\nano\meter}$, corresponding to the noise floor of the AFM tool. Further details the AFM surface roughness analysis are provided in the supporting information \cite{Supporting}.

% Orientation confirmation by XRD
\subsection{Face Orientation}
Crystallographic faces were indexed using X-ray diffraction [Fig.~\ref{diamond_surface}(f)], showing a peak at $2\theta=\SI{44}{\degree}$ that corresponds to first order Bragg diffraction from \{111\} planes.  Since the spacing between individual (111) planes is $d = a/\sqrt{3}=\SI{0.206}{\nano\meter}$, where $a = \SI{0.357}{\nano\meter}$ is the lattice constant of diamond, the first (111) diffraction peak should appear at $2\theta = 2\sin^{-1}\left(\lambda/(2d)\right) = \SI{44.0}{\degree}$. These measurements have insufficient accuracy to determine the miscut angle, but we can use the features in AFM scans like Fig.~\ref{diamond_surface}(b)-(c) to estimate the miscut based on the heights and spacings of atomically-flat plateaus.  In between Wallner lines, we typically observe atomic terraces with $\approx\SI{2}{\micro\meter}$ spacings, corresponding to a miscut angle of $\sim\SI{0.006}{\degree}$. Over larger length scales, Wallner lines of height $\approx$\SI{50}{\nano\meter} are typically separated by $\approx$\SI{10}{\micro\meter}, corresponding to an overall miscut angle $\approx\SI{0.3}{\degree}$ \cite{Supporting}.

\section{Discussion} 
%\subsection{Proposed Cleaving Mechanism}

% Physics of graphitization and cleave generation:

% Intro
While further studies are required to fully elucidate the physics underlying the LNC process, the observation of dark laser-induced lines in Figs.~\ref{optical_image}(a), \ref{cleave_process}, and~\ref{samples} suggest that the LNC process initiates and propagates a graphitization wave with a sufficiently large diameter to locally stress the SCD beyond its fracture threshold. In effect, LNC is a mechanical cleaving process in which the laser acts as a sub-micron knife striking a kerf, with the potential for creating large-area, atomically flat faces in SCD plates and gemstones for a variety of applications.

\Wallners
% 1. Localized graphitization at the nucleation site through a combination of lower surface graphitization threshold as compared to bulk, inhomogeneity of the beam intensity due to focusing, and something about texturing the surface for some reason.
Pulsed-laser graphitization and laser-guided graphitization wave propagation in SCD have been studied extensively in the context of 3D-laser writing \cite{Konov_OED_13}. In LNC, graphitization waves are propagated by focusing the laser through the starting SCD plate such that the beam fluence is sufficiently large at the bottom surface nucleation site to initiate graphitization, while the defocused beam throughout the bulk of the SCD does not create any material damage, as illustrated in Fig.~\ref{fluence}. Once the graphitization wave is initiated at the nucleation site, the ablation threshold drops abruptly such that the fluence required to propagate the wave is much lower than the bulk graphitization threshold \cite{Konov_OED_13}, allowing for the wavefront to be guided along a (111) plane by laterally scanning the laser in the $x$-direction as indicated in Fig.~\ref{cleave_process}. Using the appropriate laser fluence, spot size, and scan speed, we are able to create a graphite wave with a sufficiently large transverse diameter to fracture SCD as it propagates in a (111) plane.

% 2. Propagation of a graphite wave with dimensions that are large enough to surpass the fracture limit
Local crack formation from laser-guided graphitization waves has been previously observed in diamond \cite{Kononenko_DRM_11}, and has been attributed to the substantially greater specific volume (lower density) of graphite as compared to SCD, resulting in strong tensile stresses in the vicinity of the graphitized region. Assuming that the laser can provide adequate fluence, the requirement for producing sufficiently large internal stresses to cause cracking is determined by the area and speed of graphite generation. Romano \textit{et al.} developed an empirical threshold for crack propagation of $V\pi d^2/4R < \SI{20}{\micro\meter^3}$ per laser shot, where $V$ is the laser scan speed, $d$ is the laser spot size, and $R$ is the pulse repetition rate \cite{Konov_OED_13}. The laser micromachining system used for our LNC process has a minimum spot size $d \lesssim\SI{17}{\micro\meter}$ and a scan speed of $V = \SI{8}{\centi\meter\per\second}$ with a laser pulse repetition rate of \SI{150}{\kilo\hertz}, which gives a value of $V\pi d^2/4R \approx \SI{120}{\micro\meter^3}$ per laser shot. The fact that our processing parameters are well above the threshold for exceeding the tensile strength of SCD provides further evidence that laser-guided graphitization waves are the dominant physcial mechanism behind LNC.  Since SCD preferentially cleaves along \{111\} planes \cite{Hird_OED_13} due to the tensile strength being lowest along $\langle 111\rangle$ directions \cite{Telling_PRL_00}, the graphitization wave and resultant cleave are self guiding along the chosen (111) surface, even though the laser focus is fixed below the sample throughout the entire scan.

%\textbf{``The formation of such structures is associated with a specific challenge, and is the result of a substantially lower density of the graphitized material as compared to diamond.  Indeed, decreases in the material density under phase transition causes an expansion of the modified material and the appearance of strong tensile stresses in the surrounding diamond.  These stresses increase with cross-sectional area of the structure, and can even reach the fracture limit. In experiments conducted to date [66], the maximum cross-sectional area for these compact structures has been restricted by a value of $\approx\SI{100}{\micro\meter^2}$, to avoid any visible cracking of the adjacent diamond material.'' -\cite{Konov_OED_13}, pg. 417}

% Wallner lines/mechanical cleaving:
% 1. Wallner lines appear in mechanically cleaved diamond
% 2. They're due to a mismatch between crack velocity and SAW velocity
% 3. SAW velocity is very very high in diamond; Wallner lines are a big problem for mechanical cleaves
% 4. Velocity matching to reduce Wallner lines has been achieved in silicon.  Maybe we can do it in diamond?  Maybe we can do some sort of quasi-phase matching between the laser translation speed and SAW wave?  Maybe the graphitization wave velocity is locally much much faster than the laser translation speed since it depends on balistic hot electron transport in graphite?  Who knows, there's a lot of things we can make up here.
% 5. Cleavage along other planes by using graphite waves to engineer strain?
In contrast to traditional mechanical cleaving of diamond, LNC provides the opportunity for precise control of the cleave propagation by tuning the scan speed of the laser. Similar to what has been done with silicon stealth dicing \cite{Ohmura_JAMDSM_08}, where the laser power and stage translation speed have been tuned to achieve Wallner-line-free cleaves \cite{Casquel_ASS_11}, further analysis of the mechanisms involved in LNC should allow for drastic reduction of the Wallner lines that form during crack propagation~\cite{Hird_OED_13}. 
% While the SAW velocity in SCD is five times larger than silicon ($>\SI{10}{\kilo\meter\per\second}$ \cite{Hird_OED_13} for SCD as compared to $>\SI{2}{\kilo\meter\per\second}$ for silicon \cite{Hauch_PRL_99}) the flexibility of LNC provides opportunities for reducing Wallner lines in a similar manner. % Probably this extra information is not crucial?
Furthermore, by simultaneously scanning the laser position and focus, it may be possible to propagate cleaves along \{100\} or \{110\} planes in SCD. 

The precision and flexibility of the LNC process will make it useful for many applications. Since kerf loss is minimized, LNC may be desirable for faceting diamond gems or cutting tools. The large, perfectly flat (111) atomic planes are believed to be ideal for stabilizing the spin and electronic properties of sub-surface quantum sensors for nanoscale magnetic resonance imaging \cite{Chou2017}.
Additionally, the LNC process has potential for reducing or completely eliminating plate wedge, which adversely affects diamond photonic devices \cite{Latawiec_Optica_15,Hausmann_NP_14,Hausmann_NL_13,Gao_JLT_16}. In addition to generating (111)-faced SCD substrates for diamond electronics, photonics, and quantum technologies, LNC will also prove useful for fabricating seed crystals required for improving CVD growth of customized [111]-oriented samples \cite{Friel_DRM_09, Gracio_JPD_10,Tallaire_DRM_14,Tokuda_NAD_15,Widmann_DRM_16,Ohno2012}.

%\subsection{Process Improvements}

\section{Conclusion}
We have demonstrated a new process for fabricating (111)-faced SCD plates by laser-nucleated cleaving.  Multiple plates with thicknesses down to \SI{170}{\micro\meter} and face areas up to \SI{3.0}{\milli\meter}$\times$\SI{0.4}{\milli\meter} are reliably produced from (100) SG crystals with starting dimensions \SI{3.0}{\milli\meter}$\times$\SI{3.0}{\milli\meter}$\times$\SI{0.3}{\milli\meter}.  The plates exhibit micron-scale atomically-flat terraces, with large areas (up to \SI{1.0}{\milli\meter}$\times$\SI{0.4}{\milli\meter}) unmarred by laser damage. With the ability to produce large (111) faces out of commercially available CVD diamond plates comes the opportunity for further exploration of, and access to, SCD as an integral component in high-power electronics, architectures for quantum computing and quantum communication, and ultra-sensitive magnetometers. 

\section*{Acknowledgements}
We thank S. Szewczyk for assistance with X-ray diffraction measurements; B. Peterson and E. Johnston for assistance with micromaching lasers; and A. L. Exharos, T.-Y. Huang, and J. Tabeling for their insightful discussions and comments. This work was supported by an NSF CAREER grant (ECCS-1553511) and the Singh Center for Nanotechnology at the University of Pennsylvania.

\section*{References}

\bibliography{Parks_DRM_17}

\begin{thebibliography}{10}
\expandafter\ifx\csname url\endcsname\relax
  \def\url#1{\texttt{#1}}\fi
\expandafter\ifx\csname urlprefix\endcsname\relax\def\urlprefix{URL }\fi
\expandafter\ifx\csname href\endcsname\relax
  \def\href#1#2{#2} \def\path#1{#1}\fi

\bibitem{Balmer_JPCM_09}
R.~S. Balmer, J.~R. Brandon, S.~L. Clewes, H.~K. Dhillon, J.~M. Dodson,
  I.~Friel, P.~N. Inglis, T.~D. Madgwick, M.~L. Markham, T.~P. Mollart,
  N.~Perkins, G.~A. Scarsbrook, D.~J. Twitchen, A.~J. Whitehead, J.~J. Wilman,
  S.~M. Woollard, Chemical vapour deposition synthetic diamond: materials,
  technology and applications, Journal of Physics: Condensed Matter 21~(36)
  (2009) 364221.
\newblock \href {http://dx.doi.org/10.1088/0953-8984/21/36/364221}
  {\path{doi:10.1088/0953-8984/21/36/364221}}.

\bibitem{Gracio_JPD_10}
J.~Gracio, Q.~Fan, J.~Madaleno, Diamond growth by chemical vapour deposition,
  Journal of Physics D: Applied Physics 43~(37) (2010) 374017.
\newblock \href {http://dx.doi.org/10.1088/0022-3727/43/37/374017}
  {\path{doi:10.1088/0022-3727/43/37/374017}}.

\bibitem{E6_handbook_17}
The {Element} {Six} {CVD} diamond handbook,
  \url{https://e6cvd.com/media/wysiwyg/pdf/E6_CVD_Diamond_Handbook_A5_v10X.pdf}
  (\today).

\bibitem{Tokuda_NAD_15}
N.~Tokuda, Homoepitaxial Diamond Growth by Plasma-Enhanced Chemical Vapor
  Deposition, Springer International Publishing, 2015, pp. 1--29.
\newblock \href {http://dx.doi.org/10.1007/978-3-319-09834-0_1}
  {\path{doi:10.1007/978-3-319-09834-0_1}}.

\bibitem{Achard_DRM_11}
J.~Achard, F.~Silva, R.~Issaoui, O.~Brinza, A.~Tallaire, H.~Schneider,
  K.~Isoird, H.~Ding, S.~Kon{\'e}, M.~Pinault, et~al., Thick boron doped
  diamond single crystals for high power electronics, Diamond and Related
  Materials 20~(2) (2011) 145--152.
\newblock \href {http://dx.doi.org/10.1016/j.diamond.2010.11.014}
  {\path{doi:10.1016/j.diamond.2010.11.014}}.

\bibitem{Makino_APL_11}
T.~Makino, K.~Yoshino, N.~Sakai, K.~Uchida, S.~Koizumi, H.~Kato, D.~Takeuchi,
  M.~Ogura, K.~Oyama, T.~Matsumoto, et~al., Enhancement in emission efficiency
  of diamond deep-ultraviolet light emitting diode, Appl. Phys. Lett. 99~(6)
  (2011) 061110.
\newblock \href {http://dx.doi.org/10.1063/1.3625943}
  {\path{doi:10.1063/1.3625943}}.

\bibitem{Balducci_APL_05}
A.~Balducci, M.~Marinelli, E.~Milani, M.~Morgada, A.~Tucciarone,
  G.~Verona-Rinati, M.~Angelone, M.~Pillon, Extreme ultraviolet single-crystal
  diamond detectors by chemical vapor deposition, Applied Physics Letters
  86~(19) (2005) 193509.
\newblock \href {http://dx.doi.org/10.1063/1.1927709}
  {\path{doi:10.1063/1.1927709}}.

\bibitem{Mildren_OpEx_08}
R.~P. Mildren, J.~E. Butler, J.~R. Rabeau, {CVD}-diamond external cavity raman
  laser at 573 nm, Opt. Express 16~(23) (2008) 18950--18955.
\newblock \href {http://dx.doi.org/10.1364/OE.16.018950}
  {\path{doi:10.1364/OE.16.018950}}.

\bibitem{Hausmann_NP_14}
B.~Hausmann, I.~Bulu, V.~Venkataraman, P.~Deotare, M.~Lon{\v{c}}ar, Diamond
  nonlinear photonics, Nature Photon. 8~(5) (2014) 369--374.
\newblock \href {http://dx.doi.org/10.1038/nphoton.2014.72}
  {\path{doi:10.1038/nphoton.2014.72}}.

\bibitem{Latawiec_Optica_15}
P.~Latawiec, V.~Venkataraman, M.~J. Burek, B.~J. Hausmann, I.~Bulu,
  M.~Lon{\v{c}}ar, On-chip diamond raman laser, Optica 2~(11) (2015) 924--928.
\newblock \href {http://dx.doi.org/10.1364/OPTICA.2.000924}
  {\path{doi:10.1364/OPTICA.2.000924}}.

\bibitem{Toyli2013}
D.~M. Toyli, L.~C. Bassett, B.~B. Buckley, G.~Calusine, D.~D. Awschalom,
  Engineering and quantum control of single spins in semiconductors, MRS
  Bulletin 38 (2013) 139--143.
\newblock \href {http://dx.doi.org/10.1557/mrs.2013.21}
  {\path{doi:10.1557/mrs.2013.21}}.

\bibitem{Balasubramanian_COBC_14}
G.~Balasubramanian, A.~Lazariev, S.~R. Arumugam, D.-W. Duan, Nitrogen-vacancy
  color center in diamond--emerging nanoscale applications in bioimaging and
  biosensing, Current opinion in chemical biology 20 (2014) 69--77.
\newblock \href {http://dx.doi.org/10.1016/j.cbpa.2014.04.014}
  {\path{doi:10.1016/j.cbpa.2014.04.014}}.

\bibitem{Narayan_MT_11}
R.~J. Narayan, R.~D. Boehm, A.~V. Sumant, Medical applications of diamond
  particles \& surfaces, Materials today 14~(4) (2011) 154--163.
\newblock \href {http://dx.doi.org/10.1016/S1369-7021(11)70087-6}
  {\path{doi:10.1016/S1369-7021(11)70087-6}}.

\bibitem{Rondin_RPP_14}
L.~Rondin, J.~Tetienne, T.~Hingant, J.~Roch, P.~Maletinsky, V.~Jacques,
  Magnetometry with nitrogen-vacancy defects in diamond, Reports on Progress in
  Physics 77~(5) (2014) 056503.
\newblock \href {http://dx.doi.org/10.1088/0034-4885/77/5/056503}
  {\path{doi:10.1088/0034-4885/77/5/056503}}.

\bibitem{Aharonovich_NP_11}
I.~Aharonovich, A.~D. Greentree, S.~Prawer, Diamond photonics, Nature Photonics
  5~(7) (2011) 397--405.
\newblock \href {http://dx.doi.org/10.1038/nphoton.2011.54}
  {\path{doi:10.1038/nphoton.2011.54}}.

\bibitem{Aharonovich_AOM_14}
I.~Aharonovich, E.~Neu, Diamond nanophotonics, Advanced Optical Materials
  2~(10) (2014) 911--928.
\newblock \href {http://dx.doi.org/10.1002/adom.201400189}
  {\path{doi:10.1002/adom.201400189}}.

\bibitem{Schroder_JOSAB_16}
T.~Schr{\"o}der, S.~L. Mouradian, J.~Zheng, M.~E. Trusheim, M.~Walsh, E.~H.
  Chen, L.~Li, I.~Bayn, D.~Englund, Quantum nanophotonics in diamond [invited],
  JOSA B 33~(4) (2016) B65--B83.
\newblock \href {http://dx.doi.org/10.1364/JOSAB.33.000B65}
  {\path{doi:10.1364/JOSAB.33.000B65}}.

\bibitem{Grote_APLP_16}
R.~R. Grote, L.~C. Bassett, Single-mode optical waveguides on native
  high-refractive-index substrates, APL Photonics 1~(7) (2016) 071302.
\newblock \href {http://dx.doi.org/10.1063/1.4955065}
  {\path{doi:10.1063/1.4955065}}.

\bibitem{Gao_JLT_16}
F.~Gao, Z.~Huang, B.~Feigel, J.~Van~Erps, H.~Thienpont, R.~G. Beausoleil,
  N.~Vermeulen, Low-loss millimeter-length waveguides and grating couplers in
  single-crystal diamond, J. Lightwave Technol. 34~(23) (2016) 5576--5582.
\newblock \href {http://dx.doi.org/10.1109/JLT.2016.2622620}
  {\path{doi:10.1109/JLT.2016.2622620}}.

\bibitem{Wilks1972}
E.~M. Wilks, J.~Wilks, The resistance of diamond to abrasion, Journal of
  Physics D: Applied Physics 5~(10) (1972) 1902.
\newblock \href {http://dx.doi.org/10.1088/0022-3727/5/10/323}
  {\path{doi:10.1088/0022-3727/5/10/323}}.

\bibitem{Field_RPP_12}
J.~Field, The mechanical and strength properties of diamond, Reports on
  Progress in Physics 75~(12) (2012) 126505.
\newblock \href {http://dx.doi.org/10.1088/0034-4885/75/12/126505}
  {\path{doi:10.1088/0034-4885/75/12/126505}}.

\bibitem{Mildren_OED_13}
R.~P. Mildren, A.~Sabella, O.~Kitzler, D.~J. Spence, A.~M. McKay, Diamond raman
  laser design and performance, Optical Engineering of Diamond (2013)
  239--276\href {http://dx.doi.org/10.1002/9783527648603.ch8}
  {\path{doi:10.1002/9783527648603.ch8}}.

\bibitem{Sabella_OL_10}
A.~Sabella, J.~A. Piper, R.~P. Mildren, 1240 nm diamond raman laser operating
  near the quantum limit, Opt. Lett. 35~(23) (2010) 3874--3876.
\newblock \href {http://dx.doi.org/10.1364/OL.35.003874}
  {\path{doi:10.1364/OL.35.003874}}.

\bibitem{Katagiri_APL_04}
M.~Katagiri, J.~Isoya, S.~Koizumi, H.~Kanda, Lightly phosphorus-doped
  homoepitaxial diamond films grown by chemical vapor deposition, Appl. Phys.
  Lett. 85~(26) (2004) 6365--6367.
\newblock \href {http://dx.doi.org/10.1063/1.1840119}
  {\path{doi:10.1063/1.1840119}}.

\bibitem{Kato_APL_05}
H.~Kato, S.~Yamasaki, H.~Okushi, n-type doping of (001)-oriented
  single-crystalline diamond by phosphorus, Appl. Phys. Lett. 86~(22) (2005)
  222111.
\newblock \href {http://dx.doi.org/10.1063/1.1944228}
  {\path{doi:10.1063/1.1944228}}.

\bibitem{Kato_DRM_11}
H.~Kato, J.~Barjon, N.~Habka, T.~Matsumoto, D.~Takeuchi, H.~Okushi,
  S.~Yamasaki, Energy level of compensator states in (001) phosphorus-doped
  diamond, Diamond and Related Materials 20~(7) (2011) 1016--1019.
\newblock \href {http://dx.doi.org/10.1016/j.diamond.2011.05.021}
  {\path{doi:10.1016/j.diamond.2011.05.021}}.

\bibitem{Kato_APL_16}
H.~Kato, M.~Ogura, T.~Makino, D.~Takeuchi, S.~Yamasaki, N-type control of
  single-crystal diamond films by ultra-lightly phosphorus doping, Applied
  Physics Letters 109~(14) (2016) 142102.
\newblock \href {http://dx.doi.org/10.1063/1.4964382}
  {\path{doi:10.1063/1.4964382}}.

\bibitem{Tallaire_DRM_14}
A.~Tallaire, J.~Achard, A.~Boussadi, O.~Brinza, A.~Gicquel, I.~Kupriyanov,
  Y.~Palyanov, G.~Sakr, J.~Barjon, High quality thick {CVD} diamond films
  homoepitaxially grown on (111)-oriented substrates, Diamond and Related
  Materials 41 (2014) 34--40.
\newblock \href {http://dx.doi.org/10.1016/j.diamond.2013.11.002}
  {\path{doi:10.1016/j.diamond.2013.11.002}}.

\bibitem{Awschalom2013}
D.~D. Awschalom, L.~C. Bassett, A.~S. Dzurak, E.~L. Hu, J.~R. Petta, Quantum
  spintronics: engineering and manipulating atom-like spins in semiconductors,
  Science 339~(6124) (2013) 1174--1179.
\newblock \href {http://dx.doi.org/10.1126/science.1231364}
  {\path{doi:10.1126/science.1231364}}.

\bibitem{Hepp_PRL_14}
C.~Hepp, T.~M\"uller, V.~Waselowski, J.~N. Becker, B.~Pingault,
  H.~Sternschulte, D.~Steinm\"uller-Nethl, A.~Gali, J.~R. Maze, M.~Atat\"ure,
  C.~Becher, Electronic structure of the silicon vacancy color center in
  diamond, Phys. Rev. Lett. 112 (2014) 036405.
\newblock \href {http://dx.doi.org/10.1103/PhysRevLett.112.036405}
  {\path{doi:10.1103/PhysRevLett.112.036405}}.

\bibitem{Bhaskar_arXiv_16}
M.~K. Bhaskar, D.~D. Sukachev, A.~Sipahigil, R.~E. Evans, M.~J. Burek, C.~T.
  Nguyen, L.~J. Rogers, P.~Siyushev, M.~H. Metsch, H.~Park, F.~Jelezko,
  M.~Lon\ifmmode~\check{c}\else \v{c}\fi{}ar, M.~D. Lukin, Quantum nonlinear
  optics with a germanium-vacancy color center in a nanoscale diamond
  waveguide, Phys. Rev. Lett. 118 (2017) 223603.
\newblock \href {http://dx.doi.org/10.1103/PhysRevLett.118.223603}
  {\path{doi:10.1103/PhysRevLett.118.223603}}.

\bibitem{Neu_APL_14}
E.~Neu, P.~Appel, M.~Ganzhorn, J.~Miguel-S{\'a}nchez, M.~Lesik, V.~Mille,
  V.~Jacques, A.~Tallaire, J.~Achard, P.~Maletinsky, Photonic nano-structures
  on (111)-oriented diamond, Appl. Phys. Lett. 104~(15) (2014) 153108.
\newblock \href {http://dx.doi.org/10.1063/1.4871580}
  {\path{doi:10.1063/1.4871580}}.

\bibitem{Bernien2013}
H.~Bernien, B.~Hensen, W.~Pfaff, G.~Koolstra, M.~S. Blok, L.~Robledo, T.~H.
  Taminiau, M.~Markham, D.~J. Twitchen, L.~Childress, R.~Hanson, Heralded
  entanglement between solid-state qubits separated by three metres, Nature
  497~(7447) (2013) 86--90.
\newblock \href {http://dx.doi.org/10.1038/nature12016}
  {\path{doi:10.1038/nature12016}}.

\bibitem{Pham_PRB_12}
L.~M. Pham, N.~Bar-Gill, D.~Le~Sage, C.~Belthangady, A.~Stacey, M.~Markham,
  D.~J. Twitchen, M.~D. Lukin, R.~L. Walsworth, Enhanced metrology using
  preferential orientation of nitrogen-vacancy centers in diamond, Phys. Rev. B
  86 (2012) 121202.
\newblock \href {http://dx.doi.org/10.1103/PhysRevB.86.121202}
  {\path{doi:10.1103/PhysRevB.86.121202}}.

\bibitem{Ozawa_APE_17}
H.~Ozawa, K.~Tahara, H.~Ishiwata, M.~Hatano, T.~Iwasaki, Formation of perfectly
  aligned nitrogen-vacancy-center ensembles in chemical-vapor-deposition-grown
  diamond (111), Appl. Phys. Express 10~(4) (2017) 045501.
\newblock \href {http://dx.doi.org/10.7567/APEX.10.045501}
  {\path{doi:10.7567/APEX.10.045501}}.

\bibitem{Friel_DRM_09}
I.~Friel, S.~Clewes, H.~Dhillon, N.~Perkins, D.~Twitchen, G.~Scarsbrook,
  Control of surface and bulk crystalline quality in single crystal diamond
  grown by chemical vapour deposition, Diamond and Related Materials 18~(5–8)
  (2009) 808 -- 815, proceedings of Diamond 2008, the 19th European Conference
  on Diamond, Diamond-Like Materials, Carbon Nanotubes, Nitrides and Silicon
  Carbide.
\newblock \href {http://dx.doi.org/10.1016/j.diamond.2009.01.013}
  {\path{doi:10.1016/j.diamond.2009.01.013}}.

\bibitem{Widmann_DRM_16}
C.~Widmann, M.~Hetzl, S.~Drieschner, C.~Nebel, Homoepitaxial growth of high
  quality (111)-oriented single crystalline diamond, Diamond and Related
  Materials 72 (2017) 41 -- 46.
\newblock \href {http://dx.doi.org/10.1016/j.diamond.2016.12.020}
  {\path{doi:10.1016/j.diamond.2016.12.020}}.

\bibitem{Stoupin2016}
S.~Stoupin, S.~Antipov, J.~E. Butler, A.~V. Kolyadin, A.~Katrusha,
  {Large-surface-area diamond (111) crystal plates for applications in
  high-heat-load wavefront-preserving X-ray crystal optics}, Journal of
  Synchrotron Radiation 23~(5) (2016) 1118--1123.
\newblock \href {http://dx.doi.org/10.1107/S1600577516011796}
  {\path{doi:10.1107/S1600577516011796}}.

\bibitem{Hird_OED_13}
J.~R. Hird, Polishing and shaping of monocrystalline diamond (2013)
  71--107\href {http://dx.doi.org/10.1002/9783527648603.ch3}
  {\path{doi:10.1002/9783527648603.ch3}}.

\bibitem{Konov_OED_13}
V.~I. Konov, T.~V. Kononenko, V.~V. Kononenko, Laser Micro-and Nanoprocessing
  of Diamond Materials, Wiley Online Library, 2013, pp. 385--443.
\newblock \href {http://dx.doi.org/10.1002/9783527648603.ch12}
  {\path{doi:10.1002/9783527648603.ch12}}.

\bibitem{Kumagai_TSM_07}
M.~Kumagai, N.~Uchiyama, E.~Ohmura, R.~Sugiura, K.~Atsumi, K.~Fukumitsu,
  Advanced dicing technology for semiconductor wafer—stealth dicing, IEEE
  Transactions on Semiconductor Manufacturing 20~(3) (2007) 259--265.
\newblock \href {http://dx.doi.org/10.1109/TSM.2007.901849}
  {\path{doi:10.1109/TSM.2007.901849}}.

\bibitem{Supporting}
See the supporting information available online.

\bibitem{E6_website}
Element six, \url{https://e6cvd.com/us/} (\today).

\bibitem{Konov_LPR_12}
V.~I. Konov, Laser in micro and nanoprocessing of diamond materials, Laser \&
  Photonics Reviews 6~(6) (2012) 739--766.
\newblock \href {http://dx.doi.org/10.1002/lpor.201100030}
  {\path{doi:10.1002/lpor.201100030}}.

\bibitem{Rabinovitch_JAP_06}
A.~Rabinovitch, V.~Frid, D.~Bahat, Wallner lines revisited, J. Appl. Phys. 99
  (2006) 076102.
\newblock \href {http://dx.doi.org/10.1063/1.2181692}
  {\path{doi:10.1063/1.2181692}}.

\bibitem{DDK}
Applied diamond, inc., \url{http://usapplieddiamond.com/} (\today).

\bibitem{Kononenko_DRM_11}
T.~Kononenko, V.~Konov, S.~Pimenov, N.~Rossukanyi, A.~Rukovishnikov, V.~Romano,
  Three-dimensional laser writing in diamond bulk, Diamond and Related
  Materials 20~(2) (2011) 264--268.
\newblock \href {http://dx.doi.org/10.1016/j.diamond.2010.12.013}
  {\path{doi:10.1016/j.diamond.2010.12.013}}.

\bibitem{Telling_PRL_00}
R.~H. Telling, C.~J. Pickard, M.~C. Payne, J.~E. Field, Theoretical strength
  and cleavage of diamond, Phys. Rev. Lett. 84 (2000) 5160--5163.
\newblock \href {http://dx.doi.org/10.1103/PhysRevLett.84.5160}
  {\path{doi:10.1103/PhysRevLett.84.5160}}.

\bibitem{Ohmura_JAMDSM_08}
E.~Ohmura, M.~Kumagai, M.~Nakano, K.~Kuno, K.~Fukumitsu, H.~Morita, Analysis of
  processing mechanism in stealth dicing of ultra thin silicon wafer, Journal
  of Advanced Mechanical Design, Systems, and Manufacturing 2~(4) (2008)
  540--549.
\newblock \href {http://dx.doi.org/10.1299/jamdsm.2.540}
  {\path{doi:10.1299/jamdsm.2.540}}.

\bibitem{Casquel_ASS_11}
R.~Casquel, M.~Holgado, J.~Garc{\'\i}a-Ballesteros, K.~Zinoviev,
  C.~F{\'e}rnandez-S{\'a}nchez, F.~Sanza, C.~Molpeceres, M.~Laguna, A.~Llobera,
  J.~Oca{\~n}a, et~al., Uv laser-induced high resolution cleaving of {Si}
  wafers for micro--nano devices and polymeric waveguide characterization,
  Applied Surface Science 257~(12) (2011) 5424--5428.
\newblock \href {http://dx.doi.org/10.1016/j.apsusc.2010.11.021}
  {\path{doi:10.1016/j.apsusc.2010.11.021}}.

\bibitem{Chou2017}
J.-P. Chou, A.~Retzker, A.~Gali, Nitrogen-terminated diamond (111) surface for
  room-temperature quantum sensing and simulation, Nano Letters 17~(4) (2017)
  2294--2298.
\newblock \href {http://dx.doi.org/10.1021/acs.nanolett.6b05023}
  {\path{doi:10.1021/acs.nanolett.6b05023}}.

\bibitem{Hausmann_NL_13}
B.~J. Hausmann, I.~Bulu, P.~Deotare, M.~McCutcheon, V.~Venkataraman,
  M.~Markham, D.~Twitchen, M.~Lon{\v{c}}ar, Integrated high-quality factor
  optical resonators in diamond, Nano Lett. 13~(5) (2013) 1898--1902.
\newblock \href {http://dx.doi.org/10.1021/nl3037454}
  {\path{doi:10.1021/nl3037454}}.

\bibitem{Ohno2012}
K.~Ohno, F.~J. Heremans, L.~C. Bassett, B.~A. Myers, D.~M. Toyli, A.~C.~B.
  Jayich, C.~J. Palmstr{\o}m, D.~D. Awschalom, Engineering shallow spins in
  diamond with nitrogen delta-doping, Appl. Phys. Lett. 101~(8) (2012) 082413.
\newblock \href {http://dx.doi.org/10.1063/1.4748280}
  {\path{doi:10.1063/1.4748280}}.

\end{thebibliography}


\begin{thebibliography}{1}
\expandafter\ifx\csname url\endcsname\relax
  \def\url#1{\texttt{#1}}\fi
\expandafter\ifx\csname urlprefix\endcsname\relax\def\urlprefix{URL }\fi
\expandafter\ifx\csname href\endcsname\relax
  \def\href#1#2{#2} \def\path#1{#1}\fi

\bibitem{E6_website}
Element six, \url{https://e6cvd.com/us/} (\today).

\bibitem{jacobs2017}
T.~D. Jacobs, T.~Junge, L.~Pastewka, Quantitative characterization of surface
  topography using spectral analysis, Surface Topography: Metrology and
  Properties 5~(1) (2017) 013001.

\end{thebibliography}

\end{document}

\begin{frontmatter}

\title{Supporting Information for ``Fabrication of (111)-Faced Single-Crystal Diamond Plates by Laser Nucleated Cleaving''}

%% Group authors per affiliation:
\author[qel]{Samuel M. Parks}
\author[qel]{Richard R. Grote}
\author[qel,dop]{David A. Hopper}
\author[qel]{Lee C. Bassett\corref{cor}}
\ead{lbassett@seas.upenn.edu}
\address[qel]{Quantum Engineering Laboratory, Department of Electrical \& Systems Engineering,\\ University of Pennsylvania, Philadelphia, PA 19104, United States}
\address[dop]{Department of Physics, University of Pennsylvania, Philadelphia, PA 19104, United States}
\cortext[cor]{Corresponding author.}
\end{frontmatter}

\section{Sample details}
Details regarding the single-crystal diamond (SCD) starting plates as provided by the manufacturer \cite{E6_website} are provided in Table~\ref{sample_table}.  The edge and face orientations of both standard grade (SG) and electronics grade (EG) starting plates are depicted in Fig.~1(c,d) of the main text.  

\begin{table}[!ht]
\centering
\begin{tabular}{|c|c|c|}
\hline
&Electronics grade & Standard grade\\
\hline
N content&$<$~5~ppb&$<$~1~ppm\\
\hline
B content&$<$~1~ppb&$<$~0.05~ppm\\
\hline
Lateral Dims. & $\SI{2.0}{\milli\meter}\times\SI{2.0}{\milli\meter}$ & $\SI{3.0}{\milli\meter}~\times~\SI{3.0}{\milli\meter}$ \\
%\begin{tabular}[c]{@{}c@{}@{}}\SI{3.0}{\milli\meter}~x~\SI{3.0}{\milli\meter}\\ \& \\ \SI{2.6}{\milli\meter}~x~\SI{2.6}{\milli\meter} \end{tabular} \\
\hline
Thickness & \SI{0.5}{\milli\meter} & \SI{0.3}{\milli\meter} \\
\hline
%Edge orientation & \{110\} & \{100\}  \\
%\hline
Part no.&145-500-0385&  145-500-0253 \\
%\begin{tabular}[c]{@{}c@{}@{}}145-500-0253\\ \& \\ 145-500-0275-01 \end{tabular}\\
\hline
\end{tabular}\caption{\label{sample_table}Type IIa CVD-grown SCD starting plate specifications from Element6 \cite{E6_website}.}
\end{table}

% \begin{table}[!h]
% \begin{center}
% \begin{tabular}{|c|c|}
% \hline
%  \multicolumn{2}{|c|}{\textbf{Laser Parameters}}\\
%  \hline
% Wavelength & \SI{532}{\nano\meter}  \\
% \hline 
% Pulse Energy  & \SI{200}{\micro\joule}  \\
% \hline
% Repetition Rate  & \SI{150}{\kilo\hertz}  \\
% \hline
% Pulse Length & \SI{2}{\nano\sec}  \\
% \hline
% Focal Length & \SI{100}{\milli\meter}  \\
% \hline
% Spot Size & \SI{19}{\micro\meter} \\
% \hline
% Fluence & \SI{70.54}{\joule/\centi\meter^2} \\
% \hline
% \end{tabular}\caption{\label{green_params}Green IPG Micromachining laser parameters.}
% \end{center}
% \end{table}

\section{Micromachining laser details}
Specifications for the laser micromachining tools used for LNC are given in Table~\ref{laser_params}.  The \SI{198}{\nano\meter} excimer laser tool was used to pattern cleave nucleation sites on the bottom surface of double-side polished samples, while the \SI{532}{\nano\meter} Yb-doped fiber laser tool was used for initiating and propagating cleaves.  The spot size of the excimer laser is defined by a variable aperture, which is set here to a $\SI{10}{\micro\meter}\times\SI{10}{\micro\meter}$ square mask.

% DAH - removed this since we have no idea how big the beam actually is.
%whereas the \SI{532}{\nano\meter} laser is focused with a minimum beam spot size of $\SI{19}{\micro\meter}$. \note[LB]{Confirm that this is the diameter and not the radius.} \note[SP]{This is the diameter.}

\begin{table}[!h]
\centering
\begin{tabular}{|c|c|c|}
\hline
 & Excimer & Yb-doped fiber\\
 \hline
Model & IX-255 & IX-280-ML \\
 \hline
Wavelength & \SI{193}{\nano\meter} & \SI{532}{\nano\meter} \\
\hline 
Max Pulse energy  & \SI{800}{\micro\joule} & \SI{200}{\micro\joule} \\
\hline
Max Rep. rate  & \SI{120}{\hertz}  & \SI{300}{\kilo\hertz} \\
\hline
Pulse Length & \SI{10}{\nano\second} & \SI{2}{\nano\second} \\
\hline
Min Spot diam. & \SI{10}{\micro\meter} (apertured) & $<$\SI{17}{\micro\meter} \\
\hline
Max Fluence & \SI{25}{\joule\per\centi\meter^2} & \SI{90}{\joule\per\centi\meter^2} \\
%Mask Size & \SI{10}{\micro\meter}$\times$\SI{10}{\micro\meter}  \\
\hline
\end{tabular}\caption{\label{laser_params}Specifications for the micromachining lasers used (IPG Photonics).}
\end{table}
%\note[LB]{The parameters for the UV laser were wildly different from the spec sheet online. Should we confirm with the cleanroom staff?}}

\subsection{Finding and setting focus}
\begin{figure}[t!]
\centerline{\includegraphics[scale=0.65] {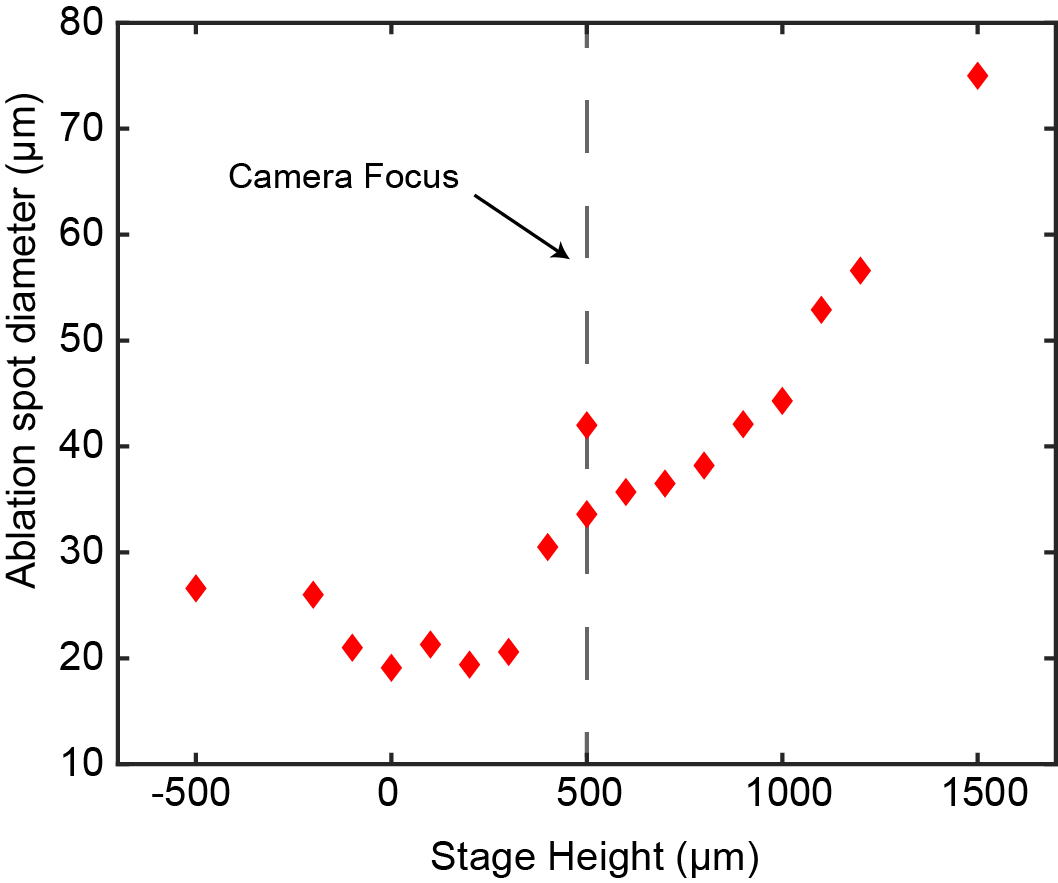}}
\caption{\label{focus}
Ablation spot diameter on the top polished diamond surface for varying stage heights. Increasing stage height corresponds to the sample surface being translated towards the laser focusing lens.}
\end{figure}

The focus of the \SI{532}{\nano\meter} laser is found by setting the laser power to 100\% with a reprate of \SI{150}{\kilo\hertz} and attempting to ablate the surface at different stage heights. The results of this calibration can be seen in Fig.~\ref{focus}. The minimum ablated spot diameter gives us an upper bound on the beam waist of \SI{17}{\micro\meter}, however the exact value if difficult to measure due to the nonlinear nature of the ablation process. The laser focus is determined to be the stage height producing the minimum ablated region width. In our coordinate system, increasing stage heights imply that the sample is being translated closer to the focusing lens. We found that the inspection camera focus and the laser focus differed by $\approx$\SI{500}{\micro\meter} and that care had to be taken to position the diamond sample in the proper location such that the top face was below the ablation threshold as is seen in Figure~3 of the main text. This calibration is performed before every cleave process to ensure that the sample is positioned properly.

\subsection{Ablation rate}
The ablation rate as a function of pulse energy in Fig.~3 of the main text was measured by setting the top diamond surface at the point of laser focus, fixing the laser rep rate at \SI{150}{\kilo\hertz}, and linearly varying the laser pulse energy between \SI{10}{\micro\joule} and \SI{200}{\micro\joule} in 10 steps with a variable attenuator to produce the series of ablation sites shown in Fig.~\ref{nuc_site}(a). Five exposures at each pulse energy were performed and averaged to arrive at the measured depth of the resulting ablation site, which were measured by AFM.

%\note[RG]{Wait, so the laser focus is set \SI{415}{\micro\meter} below the surface, but this is somehow a higher fluence than the bottom surface of a \SI{500}{\micro\meter}-thick EG plate?  I'm guessing that the distances were scaled by the refractive index of diamond, but we should be clear when discussing this.} \note[SP]{It's more accurate to say that the z-axis stage was translated upward, toward the beam objective, 415um (before that translation, the beam would be focused on the surface).}

%The fluence as a function of depth was calculated by relating the known energy per pulse of the laser (\SI{200}{\micro\joule} with no attenuation) to the diameter of the beam at a calculated depth within the SCD sample. The ablation rate could then be inferred based on the fluence at a particular depth by $\SI{200}{\micro\joule}/A$ where $A$ is beam waist cross-sectional area. When the beam is focused through the SCD diamond, the backside of the sample is closer to focus and therefore experiences a higher fluence than the sample surface where the beam is much more defocused. This allows for a selective ablation of the backside of the sample with little to no effect on the sample top surface closest to the beam objective. 

\subsection{Creating the nucleation site}
\begin{figure}[t!]
\centerline{\includegraphics[scale=1] {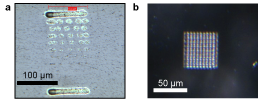}}
\caption{\label{nuc_site} 
(a) Region used for measuring ablation rate in figure 3 of the main text.
(b) Nucleation site fabricated using the UV Excimer laser [Tab. \ref{laser_params}].}
\end{figure}

The textured nucleation site shown in the main text [Fig.~2(a)] is fabricated using an Excimer laser [Tab. \ref{laser_params}]. The nucleation site is fabricated on the bottom surface of the SCD plate (see main text for surface name convention). A \SI{50}{\micro\meter}x\SI{50}{\micro\meter} area of ablated diamond [Fig.~\ref{nuc_site}(b)] is created by rastering the \SI{10}{\micro\meter}x\SI{10}{\micro\meter} aperture mask in steps of \SI{5}{\micro\meter} in two perpendicular directions. The roughened region produced is then suitable for nucleating a cleave.

\section{Processing details}

\subsection{Cleaving Single-Side Polished SG-SCD}
\begin{figure}[t!]
\includegraphics[scale=0.98] {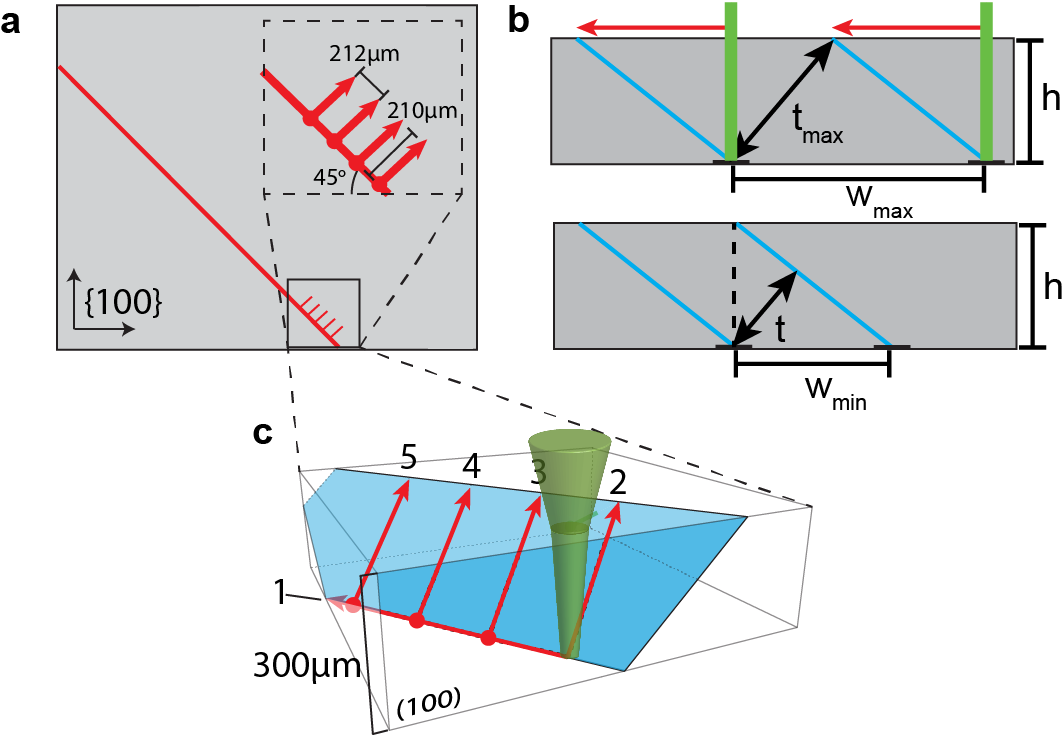}
\caption{\label{cleave_process} Illustration of the cleaving process (a) Orientation of the scribe line cut with respect to the standard orientation of SG-SCD (b) Side-on profile view of the process for cleaving a \{111\} plate with parallel faces. $w_{\textrm{max}}$ denotes the maximum lateral separation between nucleation sites, $t_{\textrm{max}}$ denotes the maximum plate thickness, $w_{\textrm{min}}$ denotes the minimum lateral separation between nucleation sites, and $t$ denotes the minimum plate thickness. (c) Expanded image of the set of cleaves required for the scribe-assisted SG-SCD cleaving process.}
\end{figure}

The unpolished-surface of single-side polished (SSP) starting plates allowed for the formation (111)-oriented SG-SCD plates without the need for a nucleation site. We found that a scribe line could be cut into the rough backside using selective ablation that allows for a cleave to propagate the length of the sample unhindered,\---\,producing (111) faces unmarred by cleave lines for $>\SI{2}{\milli\meter}$. This procedure, outlined in Figure~\ref{cleave_process}, requires that the roughened surface be placed on the bottom. A scribe line is then created on the bottom surface, oriented along the desired crystal plane, using the selective ablation procedure with the laser reprate set at \SI{75}{\kilo\hertz} (half the normal operating reprate of \SI{150}{\kilo\hertz}). Following creation of the scribe line, the laser is rastered across 3-5 lines separated by $=\SI{212}{\micro\meter}$ at 100$\%$ power [Fig.~\ref{cleave_process}(a)]. The first raster initiates the cleave, and the subsequent lines propagate the cleave across the entire scribe line to the sample edge.

%The sample is mounted with the rough side down, the green laser is focused at the surface of the SCD, and the Z axis of the sample stage is translated \SI{415}{\micro\meter} closer to the laser objective. The scribe line cut is performed at 50\% power in order to avoid unwanted cracking and is translated in the direction and to the length of the desired cleave (usually across the entire sample) along a path at a \SI{45}{\degree} angle from a \{100\} facet as seen in Fig.~\ref{cleave_process}(a) and (c). This cut allows for the cleave to continue propagating across the sample without the need to perform cleave lines continuously. After the scribe line, the laser is set to 100\% power, and 4 laser rasters of \SI{210}{\micro\meter} length, spaced by \SI{212}{\micro\meter} are performed sequentially within $\sim \SI{0.1}{\second}$ (Cleaves 2-5 in Fig.~\ref{cleave_process}(a, c)). This procedure results in a cleaved face propagating the length of the scribe line, creating a large unmarred (111) crystal face up to \SI{300}{\micro\meter} $\times$ \SI{3000}{\micro\meter}).

\subsection{Controlling plate thickness}
The (111)-faced SCD plate thickness is chosen by considering the distance between the two desired crystal faces, $t$, and then selecting the appropriate distance between nucleation sites, $w$, as shown in Fig.~\ref{cleave_process}(b). We generally observe that the laser cannot generate or propagate a cleave when intersecting a previously-created crystal face.  Hence the minimum plate thickness we can reliably produce is given by $t_{\textrm{min}}=h\cos(\SI{55}{\degree})$, where $h$ is the sample thickness. We occasionally generate thinner samples by accident, and we anticipate that by performing the cleaves on alternating sides of the diamond, thinner plates could be obtained by this method. The maximum thickness that results in two parallel, overlapping (111) faces is also shown in Fig.~\ref{cleave_process}(b).

\section{Surface topography analysis} 

\begin{figure}[t!]
\centerline{\includegraphics[scale=1] {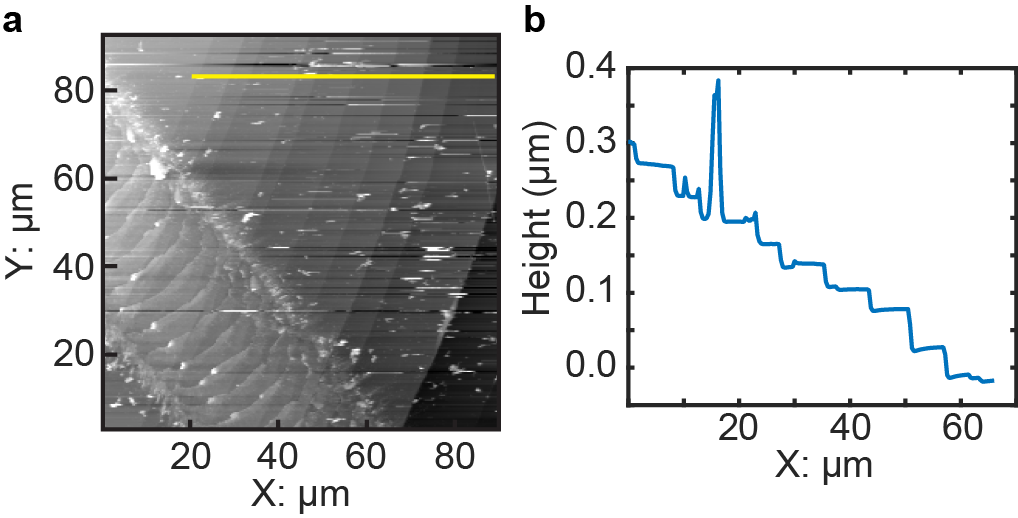}}
\caption{\label{miscut}
(a) Atomic force micrograph of cleaved (111) face. A portion of the ablated region after a cleave propagation line is seen in the bottom left corner, and a line trace is specified in yellow near the top, seen in (b). (b) Line trace of a cleaved (111) face.}
\end{figure}

\subsection{Power spectral density} Power spectral density (PSD) analysis of the surfaces of LNC fabricated (111) plates is used to characterize the surface roughness at distinct length scales of topographical features relevant for the various uses of LNC SCD (111) plates. The PSD can be integrated in the spatial frequency domain to infer roughness characteristics for a given length scale \cite{jacobs2017}. Peaks in a given PSD curve correspond to distinct frequencies of roughness, and their heights are related to their contribution to the RMS roughness at that spatial frequency.

We start with the discrete inverse and forward discrete Fourier transforms describing the surface topography, $h_x$:
\begin{eqnarray}
h_{x} = \frac{1}{L_{x}} \sum_{q}{\tilde{h}_{q}e^{iqx}} \\
\tilde{h}_{q} = l_{x} \sum_{x}{h_{x}e^{-iqx}},
\end{eqnarray}
\noindent where $L_{x}$ is the length of the AFM scan in the x-direction, $q$ is the vector of spatial frequencies, and $l_{x}$ is the pixel size of the AFM scan, $l_{x} = N_{x}/L_{x}$ where $N_{x}$ is the number of points across a single horizontal line scan. The PSD of the topography above is then defined as 
\begin{equation}\label{PSD}
C_{q}^{1D} = L_{x}^{-1} |\tilde{h_{q}}|^2
\end{equation}
It follows that Eq. \ref{PSD} is related to the RMS roughness ($R_q$) by
\begin{equation}
R_q = \sqrt{L_{x}^{-1} \sum_{q}{C_{q}^{1D}}}
\end{equation}
%By summing over the set of spatial frequencies $q$ greater than that which correspond to the spatial frequency of atomic steps, ($q = \SI{4.76e5}{\meter^{-1}}$), the components of roughness found only between atomic steps is considered. This allows for two distinct perspectives of the roughness of cleaved surfaces-- large-scale roughness that incorporates atomic step edges (summing all values of $q$) and small-scale roughness found between atomic steps (summing $q>\SI{4.919e5}{\meter^{-1}}$).
To calculate the PSD, for a given AFM scan, we use the fast Fourier transform (FFT) algorithm to calculate the Fourier components $h_q$ and corresponding PSD for each line scan of the image, and then average them together to calculate the PSD for the full image.  No image adjustments or corrections have been applied to the AFM data prior to this analysis. Note that, as calculated, the PSD only reflects spatial frequencies in the (horizontal) $x$-direction of each AFM image, so the specific shape of the curve can depend on the orientation of the image relative to specific sample features.  However, the integrated value of the RMS roughness is independent of sample orientation.

\subsection{Miscut measurement}
We estimate the miscut of LNC-fabricated (111) plates using large-scale AFM scans. 
%The accepted value for the atomic step height of a (111) oriented crystallographic plane according to the Natl. Bur. Stand. (U.S.), Circ. 539, II, 5, (1953) is \SI{0.206}{\nano\meter}. Therefore, the miscut angle across the width of a \SI{40}{\micro\meter} AFM scan is $\arctan(0.206n/40000) = \theta_{miscut}$ where $n$ is the number of atomic steps across the AFM scan. The upper bound for this value is surfaces that exhibit mostly bisteps (two stacked atomic steps) that step exclusively upward/downward across the entire AFM scan. Since the average lateral separation in atomic steps as determined using PSD analysis of surface roughness is $\approx\SI{2}{\micro\meter}$, the upper bound for the number of atomic bisteps for a given distance is 0.5 steps per micron, therefore we would expect on average 20 atomic steps across a $\SI{40}{\micro\meter}$ scan. It then follows that $\theta_{miscut} = \arctan(\SI{0.412}{\nano\meter}*20/\SI{40000}{\nano\meter})$ = \SI{0.0118}{\degree}. This method for miscut calculation was also demonstrated by N. Tokuda \cite{Tokuda_NAD_15}.
Figure~\ref{miscut} is a typical example of a region on cleaved (111) surfaces exhibiting Wallner lines. The miscut angle is calculated by counting the number and height of steps in line scans such as in Fig. \ref{miscut}(b). In those data, a distance of \SI{67}{\micro\meter} was scanned with a total vertical offset of \SI{0.31}{\micro\meter}, corresponding to a miscut angle of $\tan^{-1}(0.31/67) = \SI{0.27}{\degree}$.

\bibliography{Parks_SI}